# Three-Dimensional Dose Prediction for Lung IMRT Patients with Deep Neural Networks: Robust Learning from Heterogeneous Beam Configurations

A.M. Barragán-Montero[1, 2], D. Nguyen[1], W. Lu[1], M. Lin[1], R. Norouzi-Kandalan[1], X. Geets[2, 3], E. Sterpin[2, 4], S. Jiang[1,‡]

[1]Medical Artificial Intelligence and Automation (MAIA) Laboratory, Department of Radiation Oncology, University of Texas Southwestern Medical Center, Dallas, TX, USA

[2]Center of Molecular Imaging, Radiotherapy and Oncology (MIRO), Université catholique de Louvain, Brussels, Belgium

[3]Department of Radiation Oncology, Cliniques universitaires Saint-Luc, Brussels, Belgium

[4]Laboratory of Experimental Radiotherapy, Department of Oncology, KU Leuven, Leuven, Belgium

‡Corresponding author. E-mail: Steve.Jiang@UTSouthwestern.Edu

## ABSTRACT

**Purpose:** The use of neural networks to directly predict three-dimensional dose distributions for automatic planning is becoming popular. However, the existing methods use only patient anatomy as input and assume consistent beam configuration for all patients in the training database. The purpose of this work is to develop a more general model that considers variable beam configurations in addition to patient anatomy to achieve more comprehensive automatic planning with a potentially easier clinical implementation, without the need to train specific models for different beam settings.

**Methods:** The proposed *anatomy and beam* (AB) model is based on our newly developed deep learning architecture, hierarchically densely connected U-Net (HD U-Net), which combines U-Net and DenseNet. The AB model contains 10 input channels, one for beam setup and the other 9 for anatomical information (PTV and organs). The beam setup information is represented by a 3D matrix of the non-modulated beam's eye view ray-tracing dose distribution. We used a set of images from 129 patients with lung cancer treated with IMRT with heterogeneous beam configurations (4 to 9 beams of various orientations) for training/validation (100 patients) and testing (29 patients). Mean squared error was used as the loss function. We evaluated the model's accuracy by comparing the mean dose, maximum dose, and other relevant dose-volume metrics for the predicted dose distribution against those of the clinically delivered dose distribution. Dice similarity coefficients were computed to address the spatial correspondence of the isodose volumes between the predicted and clinically delivered doses. The model was also compared with our previous work, the *anatomy only* (AO) model, which does not consider beam setup information and uses only 9 channels for anatomical information.

**Results:** The AB model outperformed the AO model, especially in the low and medium dose regions. In terms of dose volume metrics, AB outperformed AO is about 1-2%. The largest improvement was found to be about 5% in lung volume receiving a dose of 5 Gy or more ($V_5$). The improvement for spinal cord maximum dose was also important, i.e., 3.6% for cross-validation and 2.6% for testing. The AB model achieved Dice scores for isodose volumes as much as 10% higher than the AO model in low and medium dose regions and about 2% to 5% higher in high dose regions.

**Conclusions:** The AO model, which does not use beam configuration as input, can still predict dose distributions with reasonable accuracy in high dose regions but introduces large errors in low and medium dose regions for IMRT cases with variable beam numbers and orientations. The proposed AB model outperforms the AO model substantially in low and medium dose regions and slightly in high dose regions by considering beam setup information through a cumulative non-modulated beam's eye view ray-tracing dose distribution. This new model represents a major step forward towards predicting 3D dose distributions in real clinical practice, where beam configuration could vary from patient to patient, from planner to planner, and from institution to institution.

## 1. INTRODUCTION

Current treatment planning systems for radiation therapy use advanced software to solve an inverse optimization problem,[1] which aims to determine the optimal treatment and machine parameters from an a priori specified set of dose objectives for the target and organs at risk (OARs). The fastest software can provide a solution to this problem within seconds. However, the medical physicist or dosimetrist still fine tunes the dose objectives manually until the desired dose distribution is achieved. This results in a heuristic and time-consuming process (from several hours to days), which entails a variability[2–4] in plan quality that depends on factors such as the time available to generate the plan, the institution guidelines, or the planner's skills. This variability may lead to suboptimal plans that can compromise the final treatment outcome.[5–7] Furthermore, the extended treatment planning time greatly hinders the implementation of adaptive strategies[8, 9] and may delay treatment delivery, both of which have a negative impact on tumor control and patients' quality of life.[10–13]

To overcome these problems, the research community has concentrated its efforts on reducing this manual component by automating the treatment planning process. Several groups have come up with powerful solutions that can be classified into two branches. The first branch, here referred to as *objective-based planning (OBP)*, relies on optimization algorithms that adjust pre-set objectives to achieve the established clinical goals, with well-known implementations including the in-house software Erasmus-iCycle[14] or the Auto-Planning Engine[15] commercialized by Pinnacle (Philips Radiation Oncology, Fitchburg, WI), among others.[16–19] The second branch, what is called *knowledge-based planning (KBP)*, uses a library of plans from previous patients to predict dose volume objectives for the new patient[20–23] and is best exemplified by the popular commercial solution RapidPlan (Varian Medical Systems, Palo Alto, CA). All these alternatives for automatic planning have been tested in different patient populations and anatomical sites, and they have sped up the planning process considerably (time reduction of 70-90%) for both intensity modulated radiation therapy (IMRT) and volumetric arc therapy (VMAT)[24–26] while generating high-quality plans with less human intervention.[27–31]

Even with these advancements, the OBP and KBP methods still suffer from two main drawbacks. First, they use dose volume objectives, either zero-dimensional (dose volume points) or one-dimensional (dose volume histogram, DVH), for the delineated structures. These dose volume objectives are insensitive to spatial variations of the dose within the structures delineated and blind to those structures that are not delineated. This could lead to suboptimal plans, in terms of the spatial distribution of the dose, and may require post-processing steps in which the user manually adds planning-aid structures and re-optimizes to control these spatial features. Second, both OBP and KBP strategies still require substantial human intervention to define certain parameters needed to create the model, such as the target and OAR optimization goals for OBP[14, 29, 32] or handcrafted features that serve to match the actual patient to those in the library of patients for KBP.[20, 32, 33] Including spatial dose information[34–42] and completely removing manually extracted features are necessary to achieve a more individualized and comprehensive automatic planning.

The recent evolution of deep learning methods has motivated the use of convolutional neural networks (CNN) to predict patient-specific voxel-wise dose distributions from anatomical information (i.e., contours and/or CT), either in a slice-by-slice manner (2D)[39–41, 43] or directly as a 3D matrix.[38, 42, 44] The predicted dose distribution can later be used as an objective to automatically generate a treatment plan.[37, 45] These methods completely eliminate dependence on handcrafted features by allowing the deep network to learn its own features for prediction,[38–42] and the results reported so far are very promising. However, the performance of these deep learning methods for voxel-wise dose prediction strongly depends on the database used for training, requiring users to carefully choose patients with consistent beam configurations, such as VMAT[38] or IMRT, with fixed and equally spaced beams.[39–41] This ensures an accurate dose prediction for cases with similar beam settings, but it impedes the generalization of the model to more heterogeneous beam configurations, which is crucial for IMRT treatments where the beam number and orientations could vary greatly from patient to patient and from institution to institution. As a result, the clinical implementation of automatic planning based on this type of model appears to be unfeasible, since it would require generating specific models for each individual beam arrangement.

The current models[38–42] use only anatomical information as inputs to the CNN. In this work, we investigate the value of including both anatomical and beam setup information in the network, to build a single model that is robust to variable beam configurations. This general model can realize the full potential of deep neural networks for dose prediction, bringing closer the clinical implementation of automatic planning based on this type of method.

## 2.  MATERIALS AND METHODS

## 2. A. Model architecture

The model used for dose prediction was developed in-house, and its architecture is based on the popular U-Net, published by Ronneberger et al. in 2015.[46] The U-Net is a type of CNN that belongs to the class of fully convolutional networks,[47] and it can include both local and global features from the input images to generate a pixel-wise (two-dimensional, 2D) prediction. Our group has previously used this architecture to generate 2D dose predictions for prostate patients in a slice-by-slice manner.[39] However, to avoid errors in the superior and inferior borders of the planning target volume (PTV) and OARs inherent to this 2D strategy, we developed a three-dimensional (3D) variant of the classical 2D U-Net. Since the computational load increases with the dimensionality, our group created different models to achieve an accurate and efficient 3D dose prediction. These models are described in detail elsewhere[38] and have been tested for head and neck patients. The best performance was achieved by a model that combined two recently proposed architectures: DenseNet by Huang et al. in 2017[48] and V-Net by Milletari et al. in 2016.[49] The DenseNet densely connects its convolutional layers in a feed-forward fashion, using the feature-maps of all preceding layers as inputs for the current layer. This reduces the vanishing-gradient problem, enhances feature propagation and reuse, and decreases the number of trainable parameters. The drawback of this approach is its increased memory usage, a consequence of the dense connection between layers. To maintain a reasonable RAM usage, we modified the DenseNet to skip some connections between groups of layers, following the work of Milletari et al. In addition, Huang et al.[48] found that DenseNet architectures could utilize considerably fewer trainable parameters than non-densely connected architectures, yielding better RAM usage and better generalization of the model that outweighs the greater RAM consumption of the dense connections themselves. In particular, the convolutional layers in our model are densely connected within levels of the same resolution in the U-Net, between each max pooling and up-sampling operation. We refer to each of these levels as a "hierarchy," which motivated us to name this network "Hierarchically Densely Connected U-Net" ("HD U-Net"[38]). This HD U-Net combines DenseNet's efficient feature propagation and reuse with U-Net's ability to infer the global and local image information, while maintaining a reasonable memory usage. The detailed architecture of the HD U-Net used in this study is presented in **Figure 1**, and the technical elements regarding the operations between layers have been previously described elsewhere.[38, 44] The HD U-Net uses the Adam optimization algorithm to minimize the mean squared error (MSE) between the predicted dose ($D_{pred}$) and the clinically delivered dose ($D_c$) during training, i.e., MSE = $\frac{1}{n}\sum_{i=1}^{n}(D_{pred}^i - D_c^i)^2$, where i is the index of the voxel and n is the total number of voxels. The network combines the three operations in the legend: dense convolution, dense downsampling, and U-Net upsampling. The dense convolution uses a standard convolution with the well-known Rectified Linear Unit (ReLU[50]) as its activation function, followed by a concatenation of the previous feature map. The dense downsampling uses a strided convolution and ReLU to compute a new feature map with half of the former resolution. Max pooling is applied to the previous feature map, which is then concatenated to the new feature map. The U-Net upsampling consists of upsampling, convolution, and ReLU, followed by a concatenation of the feature map on the other side of the "U." The activation function at the last output layer was also ReLU. We did not apply any regularization method during training, i.e. the dropout rate was set to zero, and there was no batch normalization.

The model proposed in this work, called the *Anatomy and Beam* (AB) model, considers both patient anatomy and beam setup information as inputs. Hence, it contains 10 input channels (**Figure 1**): one channel for beam setup information and 9 channels for anatomical information. The beam setup information is represented by an approximate 3D cumulative dose distribution (ray-tracing style, beam's eye view, and non-modulated) from all involved beams (see Section 2.D). The anatomical information comprises 3D binary matrices or masks, i.e., the prescription dose (60 Gy) for voxels inside and 0 for voxels outside the PTV, and 1 for voxels inside and 0 for voxels outside each of the 8 relevant OARs for lung treatment planning: body, heart, esophagus, spinal cord, right and left lungs, both lungs minus the target, and carina. For some patients, certain organs are not delineated, and therefore, the corresponding channel receives an empty entry.

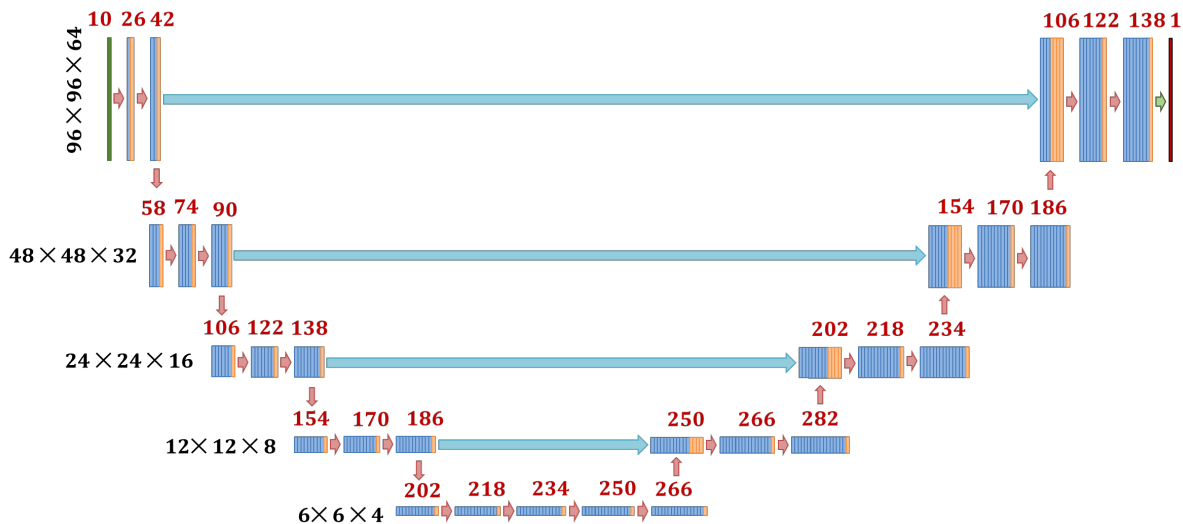

## Legend and Operations

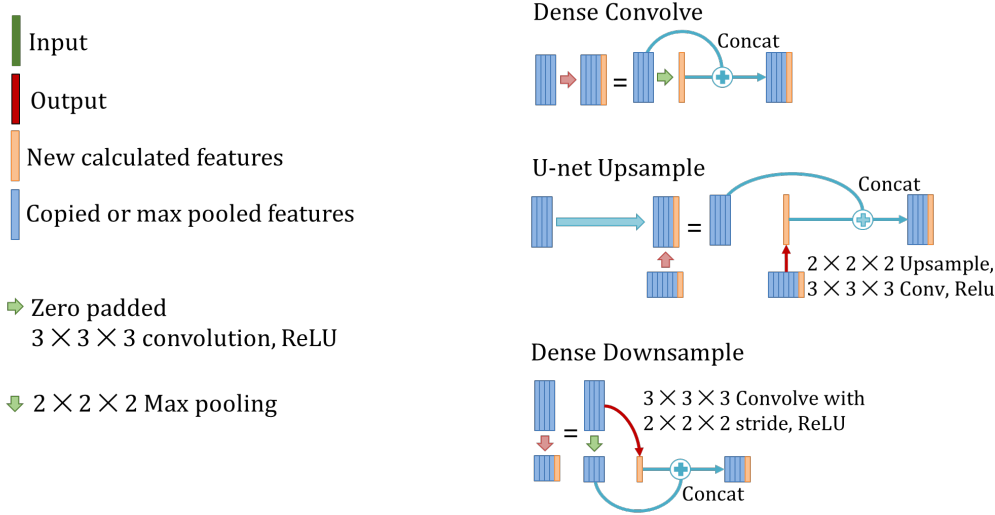

**Figure 1**. Architecture of the HD U-Net used in this study. Black numbers on the left side of the model represent the volume shape and resolution at a specific hierarchy. Red numbers represent the number of feature maps at a particular layer. Orange features represent the newly calculated features and trainable parameters to learn, while blue features are copied or max pooled features that do not need trainable parameters. The number of features (red numbers) represented here corresponds to the model and includes both anatomical and beam setup information (*AB model*, 10 input channels).

## 2. B. Patient database

The database consisted of images from 129 patients with lung cancer treated with IMRT at UT Southwestern Medical Center, which involved four different treating physicians. The database was heterogeneous in terms of number of beams (4 to 9 beams, all coplanar), beam orientation (**Figure 2** and **3**), and beam energy (6 and 10 MV). The clinically delivered dose and the contours for each patient were extracted from two different treatment planning systems: Pinnacle V8.0-V9.6 (Philips Radiation Oncology Systems, Fitchburg, WI) for patients treated before 2017 and Eclipse V13.7-V15.5 (Varian Medical Systems, Palo Alto, CA) for patients treated after 2017. All plans were created and calculated with heterogeneity correction. The target dose prescription for all patients was 60 Gy, delivered with two different fractionation protocols: 4 Gy x 15 fractions and 2 Gy x 30 fractions. The original IMRT dose and the contour masks for all patients were resampled to have a voxel resolution of 5x5x5 mm$^3$.

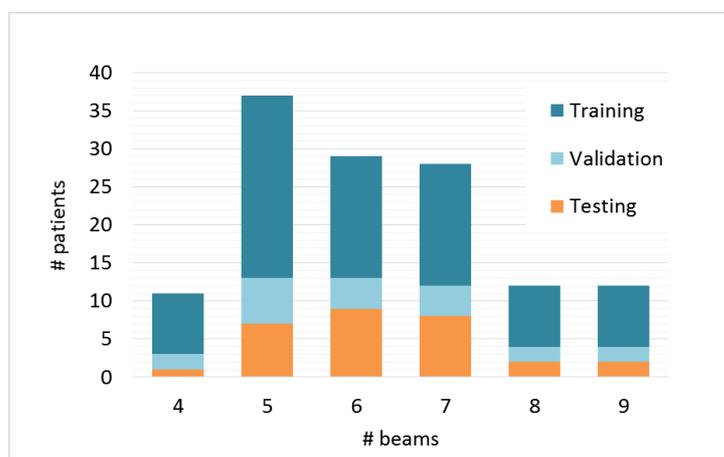

**Figure 2**. Number of patients per beam configuration (ranging from 4 to 9 beams) used for training (upper bar), validation (middle bar), and testing (lower bar).

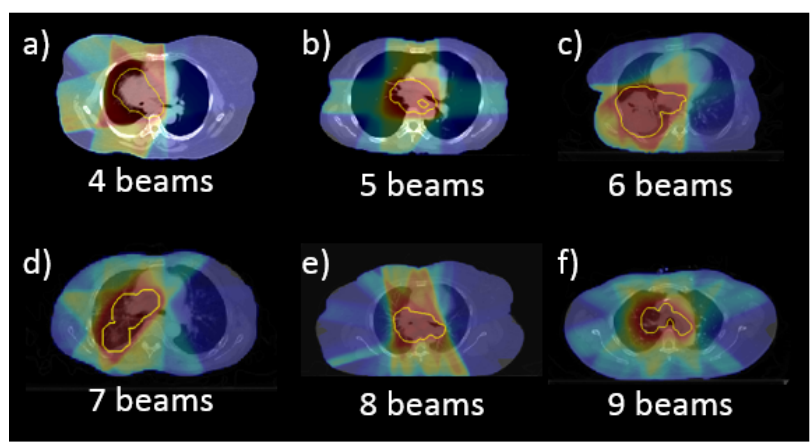

**Figure 3**. Examples of the different beam configurations from the patient database (test set). The yellow contours represent the PTV volume, and the color wash represents the dose distribution.

## 2. C. Model performance

To evaluate the model's performance and stability, we divided the database into two sets: 1) 100 patients for training and cross-validation (**Figure 2**, upper and middle bars), and 2) 29 patients for testing (**Figure 2**, lower bars). A 5-fold cross-validation procedure was applied to the 100 patient set, which was itself split into 80 patients for training and 20 patients for validation (**Figure 2**), alternating the latter along the 5 folds. When partitioning the dataset, we tried to balance the number of patients per beam configuration in the three sets: training, validation, and testing (**Figure 2**). The number of patients per beam configuration class was the same for each cross-validation fold. At every iteration, the network weights were updated based on the 80 training patients, and the loss function (MSE, see section 2.A.) was computed on the validation set. We investigated different numbers of iterations (epochs), from 100 to 1000, to find the best trade-off between optimality and training time. Since different MSE values might not directly translate into the same dose distribution quality, we also computed and analyzed relevant DVH metrics to decide the optimal number of epochs to use. Once this optimal number of epochs was found (i.e., the clinical metrics did not improve with further training), the final model was selected as the one that corresponded to the iteration with the lowest validation loss, to avoid overfitting to the training data.[51] This process as repeated for each fold, generating 5 final models that were then evaluated in the testing dataset. All these operations were performed on an NVIDIA TESLA K80 GPU with 12 GB dedicated RAM.

We also compared the *AB model* with our previous work and the current state-of-the-art, the *Anatomy Only* (*AO*) model, which contains 9 input channels for the PTV and the organs, without the beam setup information. We evaluated the accuracy of the two methods (*AB* and *AO*) by computing the average error between the predicted ($D_{p,AB}$ and $D_{p,AO}$) and clinically delivered ($D_c$) dose distributions on the mean and maximum dose values for different organs. We also analyzed the average error on relevant DVH metrics, such as the lung volume receiving a dose of at least 20 Gy ($V_{20}$) or the dose delivered to 95% of the target volume ($D_{95}$). All these values are presented as a percentage of the prescribed target dose (60 Gy). For easier comparison among patients, all doses ($D_{p,AB}$, $D_{p,AO}$, and $D_c$) were normalized to have an average dose inside the PTV equal to the prescription dose, i.e., $D_{mean}$ = 60 Gy. This normalization point serves only as a fixed point for comparison, but the user can later shift the dose to any other convenient reference, such as the $D_{95}$ of PTV equal to the prescription dose, which is often used in the clinic. We also evaluated the target dose homogeneity using the following equation for the homogeneity index: $HI = (D_2\text{-}D_{98})/D_{50}$. In addition, Dice similarity coefficients (DSC) of the isodose volumes from 5% to 95% of the prescription dose were computed for $D_{p,AB}$ and $D_{p,AO}$ and compared with those for $D_c$ to evaluate the accuracy of the spatial distribution of the doses predicted by the two models. For this purpose, three-dimensional binary masks were computed for each isodose volume containing all voxels with a dose greater than or equal to the N% of the prescription dose, in both the predicted dose (Y) and the clinically delivered doses (X). Once we had these three-dimensional binary masks (X and Y), the following operation was performed: DSC $= \frac{2|X \cap Y|}{|X|+|Y|}$.

## 2. D. Beam configuration representation

The proposed architecture (section 2.A) aims to improve the accuracy and robustness of dose prediction against a database that is heterogeneous with regard to beam arrangement. The key here is to best represent the beam configuration without greatly complicating the model architecture and, in the meantime, to provide valuable information for accurate dose distribution prediction. Ideally, a good representation should be in the dose domain and contain information about beam energy, beam aperture, and heterogeneity correction, while being computationally inexpensive. For this purpose, we use a cumulative dose distribution computed using a ray-tracing type of algorithm for all beams in the plan, without modulation, and with apertures conformal to the PTV in beam's eye view. We use a fluence-convolution broad-beam (FCBB) dose calculation method,[52, 53] which is a modified ray-tracing algorithm, involving a 2D convolution of a fluence map with a lateral spread function followed by ray-tracing based on the central axis of the beam. In our case, a dummy homogeneous fluence map (i.e., all weights equal to 1) with the aperture of the PTV projection in beam's eye view plus an isotropic margin of 5 mm is generated for each beam angle. The FCBB dose engine then uses this dummy fluence map as input, together with percentage depth dose (PDD) profiles from the Golden Beam Data (GBD)[54] provided by Varian Medical Systems, Inc. (Palo Alto, CA),[54] to compute the non-modulated dose per beam. The algorithm can generate the dose per beam in fractions of a second. Since the final computed dose per beam is given in arbitrary units, a normalization is performed after summing up all beams to make the mean dose inside the PTV equal the prescription dose. After adding up the dose corresponding to every beam, all voxels inside the PTV are overwritten to have a dose equal to the prescription dose.

Note that this study assumes that the number of beams and their orientations have been previously determined by the planner, as is commonly done in clinical practice, or will eventually be given by any beam angle optimization algorithm.

## 2. E. Additional testing

To further test the performance of our *AB model* for patients with beam configurations other than the ones included in the database described in section 2.B. (Figure 2), we used three more patients with the following beam setups: patient #1) three coplanar beams, patient #2) eleven coplanar beams, and patient #3) ten non-coplanar beams. We evaluated and compared the *AB* and *AO models* on these three patients, using the same criteria as listed in section 2.C., i.e. relevant DVH metrics, HI, and Dice coefficients for the isodose volumes.

## 3. RESULTS

The results for the average absolute error and its standard deviation (SD) on the mean and maximum dose for the target and OARs are presented in **Figure 4** for cross-validation (average prediction on the validation set for all 5 folds) and in **Figure 5** for testing (average prediction on the test set for all 5 folds). In both cases, the error on the mean and maximum doses predicted by the *AB model* was (on average) around 1% lower than on the *AO model*. The mean dose error value for the test set, averaged across all organs, was 2.28±2.01% on the *AO model* and 1.39±1.27% on the *AB model*. Likewise, the mean error on the maximum dose for the test set, averaged across all organs, was 3.97±4.73 % for the *AO model* and 2.85±3.06 % for the *AB model*. The biggest difference was found for the spinal cord maximum dose, where the *AB model*'s prediction error was up to 3.6% lower for cross-validation and 2.6% lower for testing than the *AO model*'s. **Table 1** reports some relevant DVH metrics commonly used in the clinic to evaluate lung IMRT treatments, predicted from the two models. Again, the *AB model* outperformed the *AO model*, with better prediction accuracy for all the DVH metrics considered. Although the difference in the mean average error for most metrics is rather low (around 1 - 1.5%), for other metrics such as the lung $V_{20}$ or $V_5$, the error was up to 2% and 5% lower, respectively. The spinal cord $D_2$ was also substantially lower (around 3%).

The two models predicted the dose distribution in the target volume with equivalent accuracy, with homogeneity index (HI, mean±SD) equal to 0.11±0.02 for the *AO model* and 0.08±0.02 for the *AB model*, versus 0.09±0.04 for the clinical doses, for cross-validation. Similar results were obtained for testing: HI equal to 0.10±0.03 for the *AO model* and 0.08±0.02 for the *AB model*, versus 0.09±0.03 for the clinical doses.

Dice similarity coefficients for the isodose volumes in $D_c$ versus $D_{p,AB}$ (blue) and $D_{p,AO}$ (red) are presented in **Figure 6**. The *AB model* clearly outperformed the *AO model*, with most isodose volumes having a Dice coefficient equal to or greater than 0.9, for both cross-validation and testing. In particular, the *AO model* showed a poor accuracy for the isodose volumes in the medium to low dose region, i.e., Dice < 0.9 for the isodose volumes up to 60-70% of the prescription dose, while the *AB model* achieves a Dice about 10% higher in the same region. In contrast, both models predicted the high dose region (from 80% isodose volume onwards) with comparable accuracy, though the *AB model* was still slightly superior, with Dice coefficients up to 2.5-5% higher than in the *AO model*. The lowest prediction accuracy occurs around the 40% isodose volume for both $D_{p,AB}$ and $D_{p,AO}$, but this effect appears to be much more pronounced in the *AO model*. In addition, the color-wash band representing the standard deviation of the Dice coefficient across all patients is narrower for the *AB model* for both cross-validation and testing, indicating a more stable model.

**Figure 6** also includes the Dice similarity coefficients for the isodose volumes in $D_c$ versus the input channel containing the beam representation for the *AB model*, i.e., the accumulated FCBB dose for all beams with the overwritten values inside the PTV equal to the prescription dose (Section 2.D). The FCBB dose alone seems to be an excellent approximation of $D_c$ in the low and medium dose regions, with a Dice value of about 0.9. The difference between the prediction from the *AB model* and its input channel is nearly zero up to the 20% isodose volume, at which point it starts to increase (**Figures 6.c** and **6.d.**, yellow curve). This indicates that the *AB model* uses the input channel information as it is, without further modification, and learns how to include the modulation in the dose for each beam from the 20% isodose volume onwards.

To illustrate the three-dimensional dose distribution predicted by the two models, we presented the results for one of the test patients in **Figure 7**: an axial slice located at the center of the target, as well as the corresponding DVH for $D_c$, $D_{p,AB,}$ and $D_{p,AO}$. The rest of the patients are not presented here due to limited space in the manuscript, but the behavior is similar for all of them: the *AO model* predictions show a very isotropic dose gradient that uniformly decreases from the target to the edge of the body, while the *AB model* is able to capture the dose features along the beam path due to the additional beam setup information.

The two models were trained across 150 epochs, which took about 15 hours in both cases. Additional training for a larger number of epochs was investigated but did not result in any improvement of the clinical DVH metrics under evaluation. The average prediction times and their standard deviations were 11.42 ± 0.12 s per patient for the *AO model* and 11.66 ± 0.14 s for the *AB model,* using one NVIDIA Tesla K80 card. The convergence of the two models is presented in **Figure 8**. The initial mean squared error for training and validation is much lower in the *AB model* (< 5) than in the *AO model* (> 14). This indicates that the prediction from the AB model is closer to the ground truth from the beginning, thanks to the extra input channel containing the beam setup information.

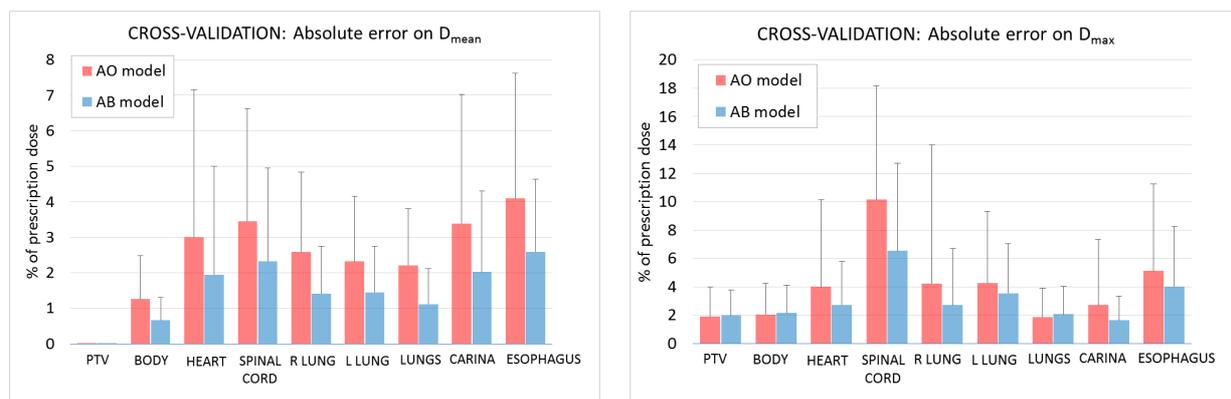

**Figure 4**. Average absolute error on the mean (left) and maximum dose (right) for the predictions ($D_{p,AB}$ and $D_{p,AO}$) versus the clinical dose ($D_c$) of all 5-fold models on the corresponding validation set for relevant organs in lung treatments. The black lines on top of the bars represent the standard deviation associated with each organ.

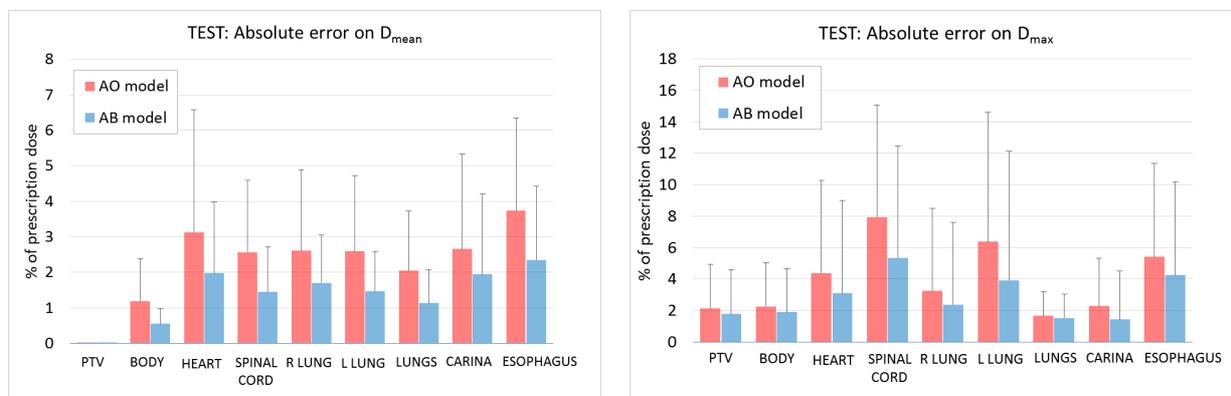

**Figure 5**. Average absolute error on the mean (left) and maximum dose (right) for the predictions ($D_{p,AB}$ and $D_{p,AO}$) versus the clinical dose ($D_c$) of all 5-fold models on the test set for relevant organs in lung treatments. The black lines on top of the bars represent the standard deviation associated with each organ.

**Table 1**. Mean absolute error and its standard deviation (mean±SD) for relevant DVH metrics on the target and on several organs for cross-validation (average prediction on the validation set for all 5 folds), and testing (average prediction on the test set for all 5 folds), for the *AO* and *AB* models. The values are expressed as percentage of the prescription dose ($D_{pre}$ = 60 Gy) for the metrics reporting the dose received by x% of volume ($D_x$), and as absolute difference for the metrics reporting the volume (in %) receiving a dose of y Gy ($V_y$).

| Mean absolute error for DVH metrics | | | | | |
|---|---|---|---|---|---|
| | | Cross-validation (mean ± SD) | | Testing (mean ± SD) | |
| | | *AO model* | *AB model* | *AO model* | *AB model* |
| **PTV** | $D_{99}$ (% of $D_{pre}$) | 3.36±3.24 | 2.70±3.06 | 3.50±2.96 | 2.54±2.56 |
| | $D_{98}$ (% of $D_{pre}$) | 2.61±2.15 | 1.95±2.09 | 2.61±2.20 | 1.71±1.73 |
| | $D_{95}$ (% of $D_{pre}$) | 1.80±1.25 | 1.10±0.86 | 1.92±1.34 | 1.08±0.96 |
| | $D_{5}$ (% of $D_{pre}$) | 0.97±0.83 | 0.81±0.75 | 1.10±1.50 | 0.94±0.74 |
| **Esophagus** | $D_{2}$ (% of $D_{pre}$) | 5.39±7.13 | 4.10±4.61 | 6.04±6.23 | 4.74±4.85 |
| | $V_{40}$ (% of volume) | 4.99±5.57 | 3.25±3.77 | 4.74±5.09 | 3.58±4.90 |
| | $V_{50}$ (% of volume) | 4.94±5.45 | 3.65±4.31 | 4.14±4.09 | 2.56±2.94 |
| **Heart** | $V_{35}$ (% of volume) | 3.40±7.78 | 2.48±6.29 | 3.32±5.41 | 2.57±4.76 |
| **Spinal cord** | $D_{2}$ (% of $D_{pre}$) | 10.10±7.82 | 6.74±6.10 | 7.64±7.12 | 5.05±4.10 |
| **Lungs** | $D_{mean}$ (% of $D_{pre}$) | 2.21±1.61 | 1.12±1.00 | 2.04±1.68 | 1.13±0.94 |
| | $V_{5}$ (% of volume) | 7.48±6.14 | 2.60±3.37 | 8.20±7.06 | 2.67±2.61 |
| | $V_{20}$ (% of volume) | 3.96±3.66 | 2.18±2.42 | 4.66±4.51 | 2.67±2.87 |

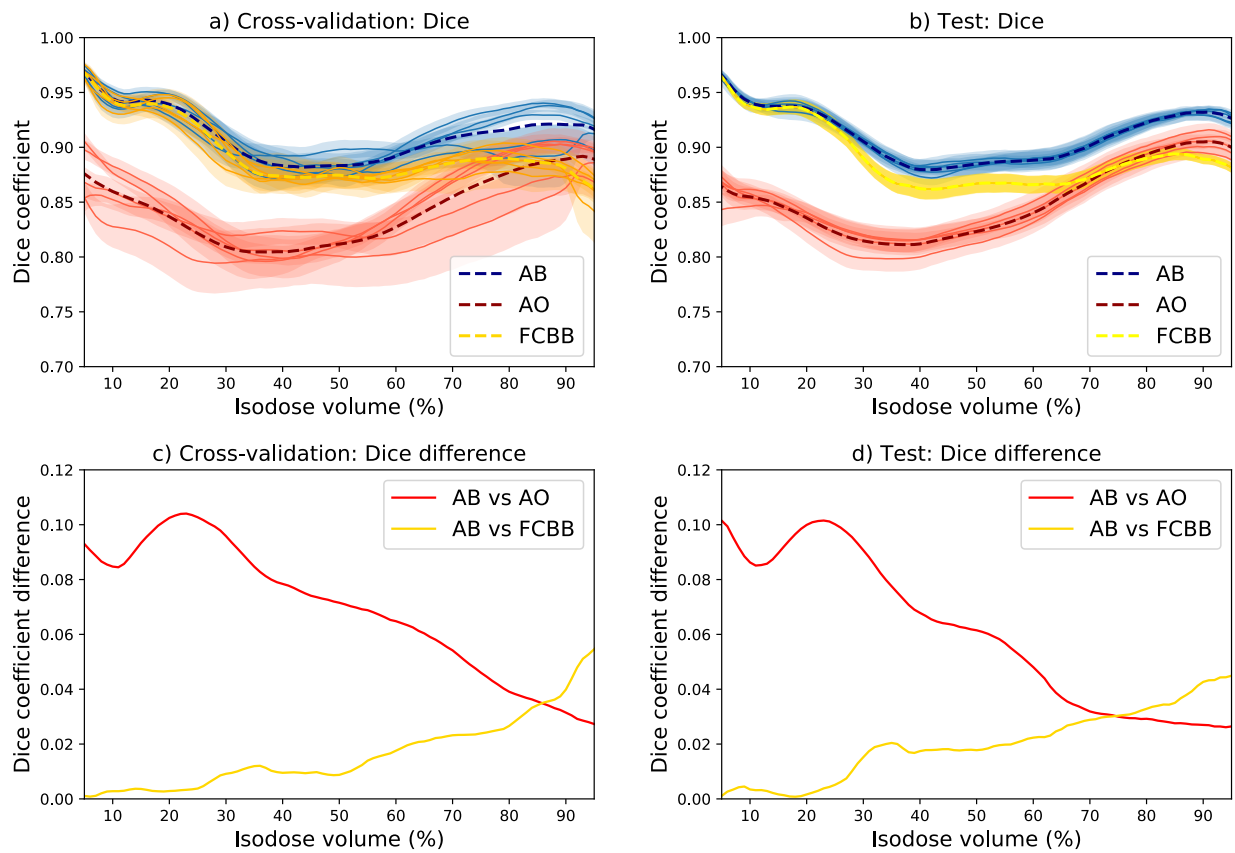

**Figure 6.** The upper plots (a. and b.) contain the Dice similarity coefficients of the isodose volumes from 5% to 95% of the prescription dose for $D_{p,AB}$ (solid blue lines), $D_{p,AO}$ (solid red lines), and the input channel of the *AB model* containing the FCBB dose (solid yellow lines), versus $D_c$, together with their corresponding average (dashed line) and standard deviation (color wash), for the 5-fold cross-validation (left) and testing (right). The bottom plots (c. and d.) contain the difference between the averaged Dice coefficient from the *AB model* versus the *AO model* (red line) and the FCBB dose used as input for the *AB model* (yellow line).

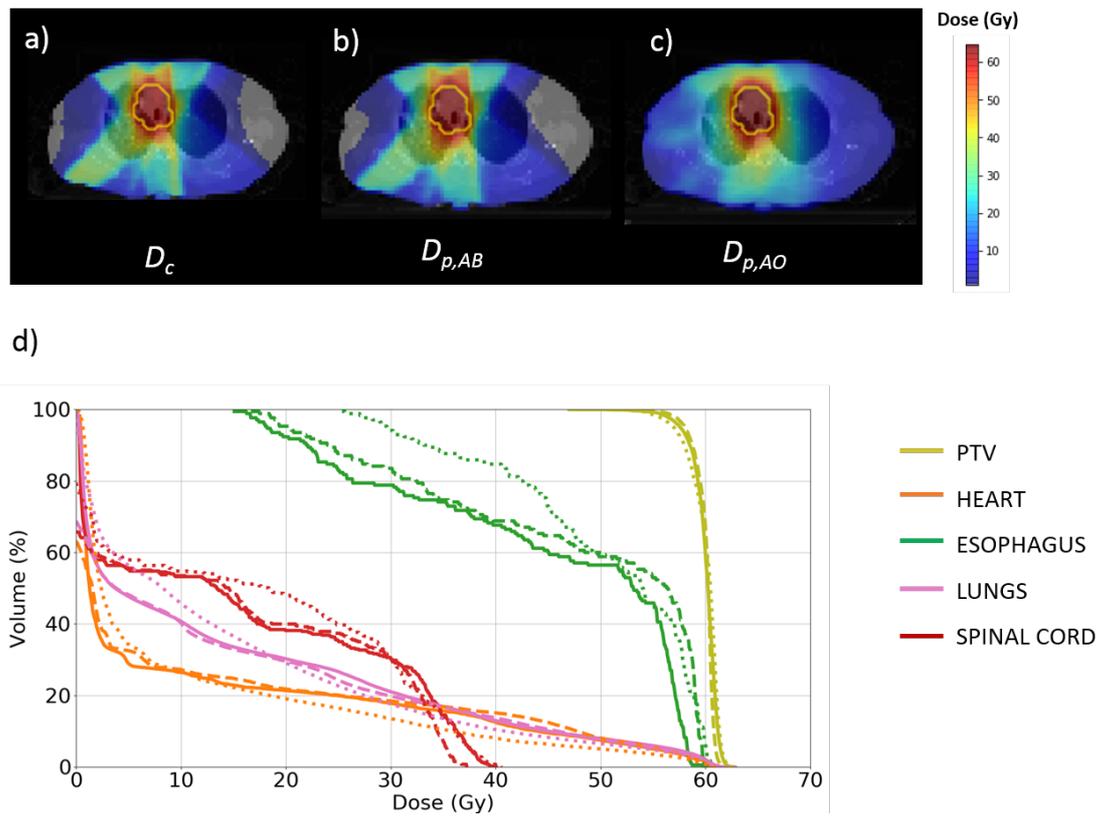

**Figure 7.** Illustration of an axial slice at the center of the target volume for one of the test patients: a) clinical dose ($D_c$), b) predicted dose from the *AB* model ($D_{p,AB}$), and c) predicted dose from the *AO* model ($D_{p,AO}$). Bottom plot (d) contains the DVHs for the three doses: solid lines correspond to the clinical dose ($D_c$), dashed lines to the prediction from *AB* model ($D_{p,AB}$), and dotted lines to the prediction from the *AO* model ($D_{p,AO}$).

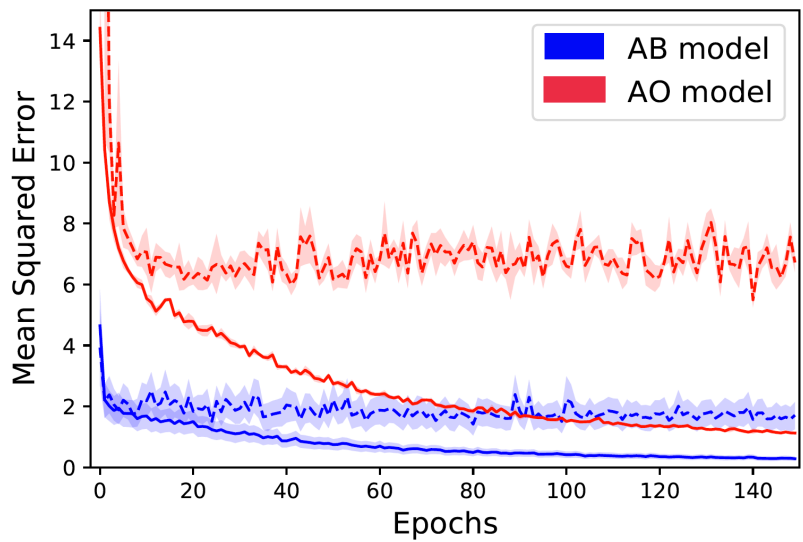

**Figure 8**. Loss function (mean squared error) evaluated for the training (solid lines) and validation sets (dashed lines), for the *AO* (red) and *AB* models (blue). The lines correspond to the average value of the loss function for all 5 models obtained after the 5-fold cross-validation process, while the color-wash bands represent the associated standard deviation.

We also tested and analyzed the *AB* and *AO models* for three additional patients with beam configurations that had not been included in the initial database used for the study (129 patients, treated with 4 to 9 beams). The absolute mean error (over all 5 folds) on relevant DVH metrics for the doses predicted with the *AB* ($D_{p,AB}$) and *AO models* ($D_{p,AO}$) for these three patients are presented in **Table 2**. The *AB model* had lower prediction errors for most metrics and outperformed the *AO model* by more than 10% in some cases. For instance, the error on the heart $V_{35}$ in patient #1 (treated with 3 coplanar beams) was 27% for the *AO model* but only 7% for the *AB model*. Similarly, the prediction error for the spinal cord $D_2$ in patient #2 (treated with 11 coplanar beams) was 16% for the *AO model* but only 4% for the *AB model*; and the error for the lungs $V_5$ in the same patient was 13% for the *AO model* and 3% for the *AB model*. The differences were less pronounced in the case of patient #3 (treated with 10 non-coplanar beams), but the *AB model* still outperformed the *AO model*'s prediction errors by 1% to 2% for most DVH metrics. These findings are confirmed by the Dice similarity coefficients on the isodose volumes, which are presented in **Figure 9**. The *AB model* outperformed the *AO model* by up to 20% in patients #1 and #2 in the low and medium dose regions, and by about 8% in patient #3. The full DVH curves for the three patients are also presented in **Figure 10**.

**Table 2**. Mean absolute error and its standard deviation (mean±SD) for relevant DVH metrics on the target and on several organs for the three patients used for additional testing (average prediction for all 5 folds), for the *AO* and *AB models*. The values are expressed as percentage of the prescription dose ($D_{pre}$ = 60 Gy) for the metrics reporting the dose received by x% of volume ($D_x$), and as absolute difference for the metrics reporting the volume (in %) receiving a dose of y Gy ($V_y$).

| | | Mean absolute error for DVH metrics | | | | | |
|---|---|---|---|---|---|---|---|
| | | P1 - 3 coplanar (mean ± SD) | | P2 - 11 coplanar (mean ± SD) | | P3 - 10 non-coplanar (mean ± SD) | |
| | | *AO model* | *AB model* | *AO model* | *AB model* | *AO model* | *AB model* |
| **PTV** | $D_{99}$ (% of $D_{pre}$) | 4.00±0.27 | 1.92±0.94 | 6.15±0.87 | 1.49±1.00 | 0.92±0.70 | 1.31±0.58 |
| | $D_{98}$ (% of $D_{pre}$) | 3.10±0.44 | 1.27±0.75 | 4.55±0.81 | 1.24±0.36 | 1.03±0.76 | 0.90±0.50 |
| | $D_{95}$ (% of $D_{pre}$) | 1.95±0.54 | 0.56±0.56 | 4.05±0.37 | 1.28±0.77 | 1.07±0.56 | 0.55±0.40 |
| | $D_5$ (% of $D_{pre}$) | 1.42±0.45 | 1.94±0.47 | 2.01±0.47 | 0.72±0.32 | 1.07±0.53 | 0.18±0.16 |
| **Esophagus** | $D_2$ (% of $D_{pre}$) | 0.73±0.24 | 0.79±0.46 | 2.54±1.26 | 1.80±1.62 | - | - |
| | $V_{40}$ (% of volume) | 1.79±1.44 | 2.50±0.51 | 0.00±0.00 | 0.00±0.00 | - | - |
| | $V_{50}$ (% of volume) | 1.67±1.25 | 2.08±0.93 | 0.00±0.00 | 0.00±0.00 | - | - |
| **Heart** | $V_{35}$ (% of volume) | 26.81±1.39 | 7.37±6.98 | 0.12±0.00 | 0.03±0.04 | 2.40±1.70 | 0.76±0.46 |
| **Spinal cord** | $D_2$ (% of $D_{pre}$) | 2.72±1.36 | 3.64±2.60 | 15.81±2.33 | 3.60±3.36 | 2.73±2.41 | 3.13±2.90 |
| **Lungs** | $D_{mean}$ (% of $D_{pre}$) | 3.05±1.69 | 1.05±0.51 | 1.83±0.16 | 0.65±0.65 | 2.26±2.27 | 0.43±0.35 |
| | $V_5$ (% of volume) | 8.57±2.52 | 2.42±0.71 | 13.26±3.60 | 2.63±2.90 | 4.92±2.55 | 1.28±0.27 |
| | $V_{20}$ (% of volume) | 5.97±2.50 | 0.85±0.86 | 0.65±0.74 | 0.49±0.48 | 4.69±6.09 | 2.24±0.59 |

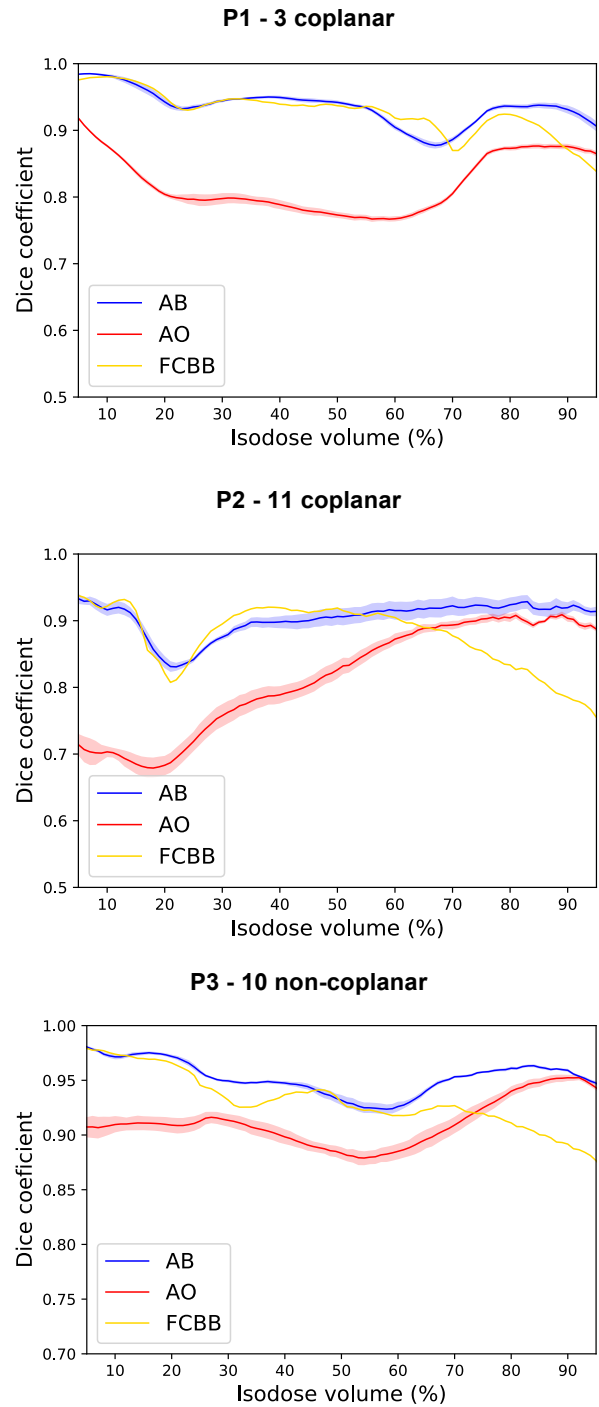

**Figure 9.** Dice similarity coefficients of the isodose volumes from 5% to 95% of the prescription dose for $D_{p,AB}$ (solid blue lines), $D_{p,AO}$ (solid red lines), and the input channel of the *AB model* containing the FCBB dose (solid yellow lines), versus $D_c$, together with their corresponding average (dashed line) and standard deviation (color wash), for the three test patients (averaged over all 5 folds).

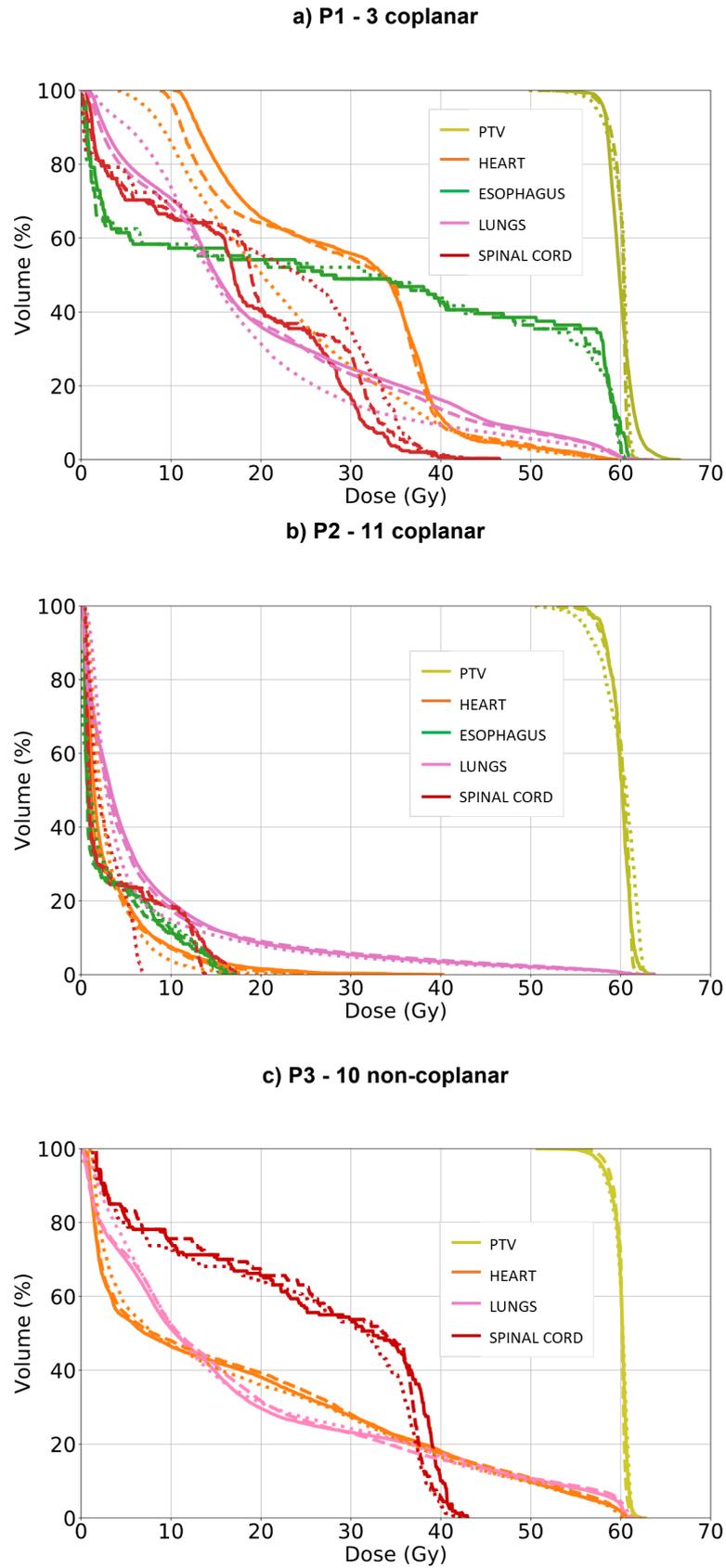

**Figure 10**. DVHs for the three patients considered (P#): solid lines correspond to the clinical dose ($D_c$), dashed lines to the prediction from *AB* model ($D_{p,AB}$), and dotted lines to the prediction from the *AO* model ($D_{p,AO}$).

## 4. DISCUSSION

The *AB model* outperformed the *AO model* in all the evaluation criteria we considered, namely, DVH metrics and Dice similarity coefficients for the isodose volumes. The difference in the prediction error between the two models was rather small in the high dose region (around 1% of the prescription dose, on average), but major differences were found for the medium to low dose regions (up to the 60-70% isodose volumes), where the beam information actually plays an important role. For these regions, the *AB model* presented a Dice coefficient 10% better than the *AO model*. Another example of the *AB model*'s superior prediction accuracy for the low dose region is the lung volume receiving at least 5 Gy ($V_5$), for which the prediction error was up to 5% lower than the *AO model*. Visual inspection of the predicted dose distributions from the two models also confirmed these results. The predictions from the *AO model* were unable to capture the dose features along the beam path, leading to a very uniform and isotropic dose fall-off. In contrast, the *AB model* accurately predicted the dose fingers corresponding to the different beam geometries.

To compare the generalization ability of the *AO* and *AB models*, we also tested them in three additional patients from our clinic, with totally new beam configurations that were not included during training. The results showed that the *AB model* achieved a prediction error more than 10% lower than the *AO model* in some cases. These findings confirm the superiority of the *AB model* to the *AO model*, even for cases where the beam configuration differs from the ones included during training.

The *AB model* was able to learn from a database that was heterogeneous in terms of beam configuration, by incorporating both anatomical and beam geometry information into the network. Our results suggest that by representing the beam configuration in the dose domain, we provide the model with valuable information about the dosimetric features that are not consistent through the database because of the variable beam arrangement. Thus, the model uses this elementary dosimetric information (FCBB) provided in the input channel and learns how to modulate it to achieve the optimal dose distribution for the given anatomy. The FCBB dose calculation used in this work is an improved ray-tracing type of algorithm, but we believe that any other elementary dose calculation algorithm can be used for the same purpose. Thus, the *AB model* represents an important step forward towards an easier and more robust implementation of automatic planning techniques, since it reduces the model's dependence on consistent beam configuration characteristics through the training patient database. This is especially true for lung IMRT treatments, where tumors occur in different positions in the thorax, and their spatial relationships with other critical organs greatly vary from patient to patient, causing more variability in beam setup than in other tumor sites, such as prostate, where the beam configuration is relatively stable. Many other types of treatment could also benefit from this improved robustness against variable beam configuration, such as IMRT - 4π treatments for brain[55] or liver,[56,57] among others.[58–60]

Regarding previous studies from other groups, the results obtained from the *AB model* (average error for the mean dose equal to 1.39±1.27%) are consistent with the values reported by McIntosh and Purdie,[37] who achieved a mean average difference of 1.33% for their lung test set using a homogeneous beam configuration (all patients treated with VMAT). They did not report the prediction error for the maximum dose, which is often more challenging to predict than the mean dose, but we consider that our model achieved excellent accuracy for this metric too, since the average for all organs was 2.85±3.06% for the test set. In addition, the prediction of the low dose region may be better thanks to the supplementary beam setup information, even for models that used the same beam configuration for all patients in the database. For instance, Moore et al.,[34] using a database where all patients (prostate and brain) were treated with VMAT, reported up to 20 Gy of difference between the predicted and real doses for regions that were far from the PTV.

In radiation therapy treatment planning, the sum of the square difference between the planned and the prescription dose is often used as the loss function.[1] This prompted us to use the MSE between the predicted and the clinically delivered doses as the loss function in our study. In addition, using MSE is computationally cheap and leads to a convex optimization problem with a stable gradient, which is easier to solve than non-convex problems. However, investigating the use of other metrics as loss functions, such as the inclusion of DVH metrics for target and organs, may be an interesting field of study.

The time needed to predict the three-dimensional dose distribution per patient was similar for the two models: *AB model* (11.66 ± 0.14 s) and *AO model* (11.42 ± 0.12 s). The time employed to compute the FCBB dose used as input to the model can be considered negligible since it was less than one second.[53] In addition, the FCBB dose per beam can later be used to feed the optimizer[52] in the treatment planning system employed to generate the plan. The predicted 3D dose matrix can then be used as a voxel-wise objective to create a treatment plan that mimics it.[35,41,43] Since the most advanced optimizers can provide a solution within seconds, the total time required to generate a full plan may be kept under a minute, if the right hardware is used. This provides a good environment to implement online adaptive strategies,[9] where the plan needs to be adapted while the patient is on the treatment couch, and every extra minute is of crucial importance. In addition, the presented model could be used as part of beam angle optimization strategies for IMRT, by generating 3D doses for different beam configurations and then selecting the optimal one according to DVH metrics or any other relevant criteria used in the clinic for plan evaluation. In this context, the generated doses could also serve as planning guidance for the dosimetrist or even as a decision support tool for the treating physician before going to treatment planning. Eventually, the dose prediction model could be used in tumor board meetings for comparison with other suitable treatments and could assist in evaluating tumor control probability and possible secondary effects. However, as is the case with every deep learning application, one must be aware of the importance of the quality of the database used for training. If the ground truth doses are suboptimal, the predicted doses will be suboptimal too, i.e., the *garbage in, garbage out* paradigm. Therefore, the medical community should encourage the construction of high quality databases created by experienced planners, which can serve to improve and standardize future clinical practice. Meanwhile, the best solution might be the use of human-assisted and dynamic workflows, where the models are trained with the existing databases (heterogeneous plan quality) and used under the supervision of physicians. The physicians will then select the highest quality results, which will be used later to update and improve the current models.

Lastly, the dose prediction models in the existing literature have been applied so far to radiation therapy treatments with photons, i.e., IMRT or VMAT. However, extending these models to proton therapy represents an extra challenge, given the sensitivity of the dose distribution to heterogeneities in the tissue traversed. In this context, providing the model with basic beam setup information along the beam path is essential, and we believe our model could be easily applied for that purpose, which we plan to study in a future work.

**5. CONCLUSION**

We used deep neural networks to build a model that can learn from a database of previous patients treated with variable beam configuration and predict the three-dimensional dose distribution for a new patient. Two models were trained and compared: the first model (AO) only accounted for the anatomy of the patient, while the second model (AB) included both anatomical and beam setup information, the latter being represented in the dose domain. The AB model showed greater accuracy and robustness against variable beam geometry than the AO model. This suggests that using a three-dimensional matrix containing elementary dose features along the beam paths as input to the model will help to achieve a more comprehensive automatic planning based on deep neural networks without the need to train specific models for every beam arrangement.

**ACKNOWLEDGEMENTS**

Ana Barragán Montero is supported by Fonds Baillet-Latour. Jonathan Feinberg edited the manuscript.

**REFERENCES**

[1] U. Oelfke and T. Bortfeld, Inverse planning for photon and proton beams, Med. Dosim. (2001).

[2] V. Batumalai, M.G. Jameson, D.F. Forstner, P. Vial, and L.C. Holloway, How important is dosimetrist experience for intensity modulated radiation therapy? A comparative analysis of a head and neck case, Pract. Radiat. Oncol. (2013).

[3] S.L. Berry, A. Boczkowski, R. Ma, J. Mechalakos, and M. Hunt, Interobserver variability in radiation therapy plan output: Results of a single-institution study, Pract. Radiat. Oncol. (2016).

[4] B.E. Nelms, G. Robinson, J. Markham, *et al.*, Variation in external beam treatment plan quality: An inter-institutional study of planners and planning systems, Pract. Radiat. Oncol. (2012).

[5] M. Marcello, M. Ebert, A. Haworth, *et al.*, Association between treatment planning and delivery factors and disease progression in prostate cancer radiotherapy: Results from the TROG 03.04 RADAR trial, Radiother. Oncol. (2018).

[6] L.J. Peters, B. O'Sullivan, J. Giralt, *et al.*, Critical impact of radiotherapy protocol compliance and quality in the treatment of advanced head and neck cancer: Results from TROG 02.02, J. Clin. Oncol. (2010).

[7] K.L. Moore, R. Schmidt, V. Moiseenko, *et al.*, Quantifying unnecessary normal tissue complication risks due to suboptimal planning: A secondary study of RTOG 0126, Int. J. Radiat. Oncol. Biol. Phys. (2015).

[8] D. Yan, F. Vicini, J. Wong, and A. Martinez, Adaptive radiation therapy, Phys. Med. Biol. (1997).

[9] S. Lim-Reinders, B.M. Keller, S. Al-Ward, A. Sahgal, and A. Kim, *Online Adaptive Radiation Therapy*, Int. J. Radiat. Oncol. Biol. Phys. (2017).

[10] J.A. González Ferreira, J. Jaén Olasolo, I. Azinovic, and B. Jeremic, *Effect of radiotherapy delay in overall treatment time on local control and survival in head and*

*neck cancer: Review of the literature*, Reports Pract. Oncol. Radiother. (2015).

11   H. Yang, W. Hu, W. Wang, P. Chen, W. Ding, and W. Luo, Replanning during intensity modulated radiation therapy improved quality of life in patients with nasopharyngeal carcinoma, Int. J. Radiat. Oncol. Biol. Phys. (2013).

12   C. A.M., Y. T., H. S., and M. A., Image-guided adaptive radiotherapy improves acute toxicity during intensity-modulated radiation therapy for head and neck cancer, J. Radiat. Oncol. (2018).

13   D.S. Møller, M.I. Holt, M. Alber, *et al.*, Adaptive radiotherapy for advanced lung cancer ensures target coverage and decreases lung dose, Radiother. Oncol. (2016).

14   S. Breedveld, P.R.M. Storchi, P.W.J. Voet, and B.J.M. Heijmen, ICycle: Integrated, multicriterial beam angle, and profile optimization for generation of coplanar and noncoplanar IMRT plans, Med. Phys. (2012).

15   I. Xhaferllari, E. Wong, K. Bzdusek, M. Lock, and J.Z. Chen, Automated IMRT planning with regional optimization using planning scripts, J. Appl. Clin. Med. Phys. (2013).

16   T.G. Purdie, R.E. Dinniwell, A. Fyles, and M.B. Sharpe, Automation and intensity modulated radiation therapy for individualized high-quality tangent breast treatment plans, Int. J. Radiat. Oncol. Biol. Phys. (2014).

17   X. Zhang, X. Li, E.M. Quan, X. Pan, and Y. Li, A methodology for automatic intensity-modulated radiation treatment planning for lung cancer, Phys. Med. Biol. (2011).

18   D.L. Craft, T.S. Hong, H.A. Shih, and T.R. Bortfeld, Improved planning time and plan quality through multicriteria optimization for intensity-modulated radiotherapy, Int. J. Radiat. Oncol. Biol. Phys. (2012).

19   S. Breedveld, D. Craft, R. van Haveren, and B. Heijmen, *Multi-criteria optimization and decision-making in radiotherapy*, Eur. J. Oper. Res. (2018).

20   B. Wu, F. Ricchetti, G. Sanguineti, *et al.*, Data-driven approach to generating achievable dose-volume histogram objectives in intensity-modulated radiotherapy planning, Int. J. Radiat. Oncol. Biol. Phys. (2011).

21   L.M. Appenzoller, J.M. Michalski, W.L. Thorstad, S. Mutic, and K.L. Moore, Predicting dose-volume histograms for organs-at-risk in IMRT planning, Med. Phys. (2012).

22   L. Yuan, Y. Ge, W.R. Lee, F.F. Yin, J.P. Kirkpatrick, and Q.J. Wu, Quantitative analysis of the factors which affect the interpatient organ-At-risk dose sparing variation in IMRT plans, Med. Phys. (2012).

23   X. Zhu, T. li, D. Thongphiew, Y. ge, F. Yin, and Q. wu, TU-E-BRB-03: A Planning Quality Evaluation Tool for Adaptive IMRT Treatment Based on Machine Learning, in *Med. Phys.*(2010).

24   J. Krayenbuehl, I. Norton, G. Studer, and M. Guckenberger, Evaluation of an automated knowledge based treatment planning system for head and neck, Radiat. Oncol. (2015).

25   A.T.Y. Chang, A.W.M. Hung, F.W.K. Cheung, *et al.*, Comparison of Planning Quality and Efficiency Between Conventional and Knowledge-based Algorithms in Nasopharyngeal Cancer Patients Using Intensity Modulated Radiation Therapy, in *Int. J. Radiat. Oncol. Biol. Phys.*(2016).

26   B. Vanderstraeten, B. Goddeeris, K. Vandecasteele, M. van Eijkeren, C. De Wagter, and Y. Lievens, Automated Instead of Manual Treatment Planning? A Plan Comparison Based on Dose-Volume Statistics and Clinical Preference, Int. J. Radiat. Oncol. Biol. Phys. (2018).

27   C.R. Hansen, A. Bertelsen, I. Hazell, *et al.*, Automatic treatment planning improves

the clinical quality of head and neck cancer treatment plans, Clin. Transl. Radiat. Oncol. (2016).

28  M. Hussein, C.P. South, M.A. Barry, *et al.*, Clinical validation and benchmarking of knowledge-based IMRT and VMAT treatment planning in pelvic anatomy, Radiother. Oncol. (2016).

29  P.W.J. Voet, M.L.P. Dirkx, S. Breedveld, A. Al-Mamgani, L. Incrocci, and B.J.M. Heijmen, Fully automated volumetric modulated arc therapy plan generation for prostate cancer patients, Int. J. Radiat. Oncol. Biol. Phys. (2014).

30  J.P. Tol, A.R. Delaney, M. Dahele, B.J. Slotman, and W.F.A.R. Verbakel, Evaluation of a knowledge-based planning solution for head and neck cancer, Int. J. Radiat. Oncol. Biol. Phys. (2015).

31  A.R. Delaney, J.P. Tol, M. Dahele, J. Cuijpers, B.J. Slotman, and W.F.A.R. Verbakel, Effect of Dosimetric Outliers on the Performance of a Commercial Knowledge-Based Planning Solution, Int. J. Radiat. Oncol. Biol. Phys. (2016).

32  B. Wu, M. Kusters, M. Kunze-busch, *et al.*, Cross-institutional knowledge-based planning (KBP) implementation and its performance comparison to Auto-Planning Engine (APE), Radiother. Oncol. (2017).

33  L. Yuan, W. Zhu, Y. Ge, *et al.*, Lung IMRT planning with automatic determination of beam angle configurations, Phys. Med. Biol. (2018).

34  S. Shiraishi and K.L. Moore, Knowledge-based prediction of three-dimensional dose distributions for external beam radiotherapy, Med. Phys. (2016).

35  C. McIntosh, M. Welch, A. McNiven, D.A. Jaffray, and T.G. Purdie, Fully automated treatment planning for head and neck radiotherapy using a voxel-based dose prediction and dose mimicking method, Phys. Med. Biol. (2017).

36  C. McIntosh and T.G. Purdie, Contextual Atlas Regression Forests: Multiple-Atlas-Based Automated Dose Prediction in Radiation Therapy, IEEE Trans. Med. Imaging (2016).

37  C. McIntosh and T.G. Purdie, Voxel-based dose prediction with multi-patient atlas selection for automated radiotherapy treatment planning, Phys. Med. Biol. (2017).

38  D. Nguyen, X. Jia, D. Sher, *et al.*, Three-Dimensional Radiotherapy Dose Prediction on Head and Neck Cancer Patients with a Hierarchically Densely Connected U-net Deep Learning Architecture, (2018).

39  D. Nguyen, T. Long, X. Jia, *et al.*, Dose Prediction with U-net: A Feasibility Study for Predicting Dose Distributions from Contours using Deep Learning on Prostate IMRT Patients, (2017).

40  X. Chen, K. Men, Y. Li, J. Yi, and J. Dai, A feasibility study on an automated method to generate patient-specific dose distributions for radiotherapy using deep learning, Med. Phys. (2018).

41  J. Fan, J. Wang, Z. Chen, C. Hu, Z. Zhang, and W. Hu, Automatic treatment planning based on three-dimensional dose distribution predicted from deep learning technique, Med. Phys. (2018).

42  V. Kearney, J.W. Chan, S. Haaf, M. Descovich, and T.D. Solberg, DoseNet : a volumetric dose prediction algorithm using 3D fully- convolutional neural networks, Phys. Med. Biol. **63**, (2018).

43  D. Nguyen, T. Long, X. Jia, *et al.*, A feasibility study for predicting optimal radiation therapy dose distributions of prostate cancer patients from patient anatomy using deep learning, Sci. Rep. (2019).

44  D. Nguyen, X. Jia, D. Sher, *et al.*, 3D radiotherapy dose prediction on head and neck cancer patients with a hierarchically densely connected U-net deep learning architecture, Phys. Med. Biol. **64**(6), 065020 (2019).

45  T. Long, M. Chen, S. Jiang, and W. Lu, Threshold-driven optimization for reference-based auto-planning, Phys. Med. Biol. (2018).

46  O. Ronneberger, P. Fischer, and T. Brox, U-net: Convolutional networks for biomedical image segmentation, in *Lect. Notes Comput. Sci. (including Subser. Lect. Notes Artif. Intell. Lect. Notes Bioinformatics)*(2015).

47  E. Shelhamer, J. Long, and T. Darrell, Fully Convolutional Networks for Semantic Segmentation, IEEE Trans. Pattern Anal. Mach. Intell. (2017).

48  G. Huang, Z. Liu, L. Van Der Maaten, and K.Q. Weinberger, Densely connected convolutional networks, in *Proc. - 30th IEEE Conf. Comput. Vis. Pattern Recognition, CVPR 2017*(2017).

49  F. Milletari, N. Navab, and S.A. Ahmadi, V-Net: Fully convolutional neural networks for volumetric medical image segmentation, in *Proc. - 2016 4th Int. Conf. 3D Vision, 3DV 2016*(2016).

50  V. Nair and G. Hinton, Rectified Linear Units Improve Restricted Boltzmann Machines, in *Proc. 27th Int. Conf. Mach. Learn.*(2010).

51  R. Caruana, S. Lawrence, and L. Giles, Overfitting in neural nets: Backpropagation, conjugate gradient, and early stopping, Nips-2000 (2000).

52  W. Lu, A non-voxel-based broad-beam (NVBB) framework for IMRT treatment planning, Phys. Med. Biol. (2010).

53  W. Lu and M. Chen, Fluence-convolution broad-beam (FCBB) dose calculation, Phys. Med. Biol. (2010).

54  G.P. Beyera, Commissioning measurements for photon beam data on three TrueBeam linear accelerators, and comparison with Trilogy and Clinac 2100 linear accelerators, J. Appl. Clin. Med. Phys. (2013).

55  V.L. Murzin, K. Woods, V. Moiseenko, *et al.*, 4π plan optimization for cortical-sparing brain radiotherapy, Radiother. Oncol. (2018).

56  D. P., L. T., R. D., *et al.*, 4PI radiation therapy for liver SBRT, Int. J. Radiat. Oncol. Biol. Phys. (2012).

57  K. Woods, D. Nguyen, A. Tran, *et al.*, Viability of Noncoplanar VMAT for liver SBRT compared with coplanar VMAT and beam orientation optimized 4π IMRT, Adv. Radiat. Oncol. (2016).

58  J.C.M. Rwigema, D. Nguyen, D.E. Heron, *et al.*, 4π noncoplanar stereotactic body radiation therapy for head-and-neck cancer: Potential to improve tumor control and late toxicity, Int. J. Radiat. Oncol. Biol. Phys. (2015).

59  D. Nguyen, J.C.M. Rwigema, V.Y. Yu, *et al.*, Feasibility of extreme dose escalation for glioblastoma multiforme using 4π radiotherapy, Radiat. Oncol. (2014).

60  A. Tran, J. Zhang, K. Woods, *et al.*, Treatment planning comparison of IMPT, VMAT and 4Π radiotherapy for prostate cases, Radiat. Oncol. (2017).